\documentclass[twocolumn, showpacs,preprintnumbers,amsmath,amssymb]{revtex4}
\usepackage{amssymb}
\usepackage{xcolor}
\usepackage{graphicx}
\usepackage{dcolumn}
\usepackage{bm}
\usepackage{multirow}
\usepackage{booktabs}
\usepackage{natbib}
\usepackage{soul}
\usepackage{amssymb}
\usepackage{amsmath}	
\usepackage{color}
\usepackage{cases} 

\usepackage{array}
\newcommand{\Kmat}{\mbox{\boldmath $\mathcal{K}$}}

\newcommand{\Vmat}{\mbox{\boldmath $\mathcal{V}$}}
\newcommand{\V}{\mbox{\boldmath $V$}}

\newcommand{\Omat}{{\bf O}}
\newcommand{\Umat}{\mbox{\boldmath $\mathcal{U}$}}
\newcommand{\etamat}{\mbox{\boldmath $\eta$}}
\newcommand{\numat}{\mbox{\boldmath $\nu$}}
\newcommand{\Cmat}{\mbox{\boldmath $\cal C$}}
\newcommand{\Smat}{\mbox{\boldmath $\cal S$}}
\newcommand{\SSmat}{\mbox{ $S$}}
\newcommand{\Xmat}{\mbox{\boldmath $X$}}

\newcommand{\MK}{\mathcal{K}}
\newcommand{\MV}{\mathcal{V}}
\newcommand{\Cmatt}{\mbox{\boldmath $\mathcal{C}$}}
\newcommand{\Smatt}{\mbox{\boldmath $\mathcal{S}$}}

\begin{document}
\preprint{APS/123-QED}

\title{Dissociative recombination of the CH$^+$ molecular ion at low energy}

\author{K. Chakrabarti$^{1,2}$}
\author{J. Zs Mezei$^{1,3,4}$}
\author{O. Motapon$^{5}$}
\author{A. Faure$^{6}$}
\author{O. Dulieu$^{7}$}
\author{K. Hassouni$^{3}$}
\author{I. F. Schneider$^{1,7}$}\email[]{ioan.schneider@univ-lehavre.fr}
\affiliation{$^{1}$LOMC-UMR6294, CNRS, Universit\'e Le Havre Normandie, 76600 Le Havre, France}%
\affiliation{$^{2}$Dept. of Mathematics, Scottish Church College, 700006 Kolkata, India}
\affiliation{$^{3}$LSPM CNRS-UPR3407, Universit{\'{e}} Paris 13, 93430 Villetaneuse, France}
\affiliation{$^{4}$HUN-REN Institute for Nuclear Research (ATOMKI), H-4001 Debrecen, Hungary}%
\affiliation{$^{5}$LPF, University of Douala, 24157 Douala, Cameroon}
\affiliation{$^{6}$IPAG, CNRS-INSU-UMR5274, Universit\'e UJF-Grenoble 1, 38400 Grenoble, France}
\affiliation{$^{7}$LAC-UMR9188, CNRS Universit\'e Paris-Saclay, F-91405 Orsay, France}%
\date{\today}

\begin{abstract}
The reactive collisions of the CH$^+$ molecular ion with electrons is studied in the framework of the multichannel quantum defect theory, taking into account the contribution of the core-excited Rydberg states. In addition to the $X ^1\Sigma^+$ ground state of the ion, we also consider the contribution to the dynamics of the $a ^3\Pi$ and $A ^1\Pi$ excited states of CH$^+$. Our results - in the case of the dissociative recombination in good agreement with the storage ring measurements -  rely on decisive improvements  - complete account of the ionisation channels and accurate evaluation of the reaction matrix - of a previously used model.
\end{abstract}


\maketitle

The hydrocarbon molecular ions are major constituents in many low temperature ionised environments such as diffuse interstellar clouds, planetary atmospheres and technological plasmas. The reactive collision of these ions with electrons play an important role governing their chemistry.

The simplest among them, CH$^+$, was first found in the interstellar molecular clouds in 1941 by Douglas and Herzberg \cite{dh1941}. Since then, its absorption lines have been observed towards many background stars, demonstrating the omnipresence of this simple carbon hydride in the diffuse interstellar media (ISM). The mechanism by which it forms remains however still puzzling. 
It is believed to be formed by the reaction:
\begin{equation}\label{eq:CHform}
\mbox{C}^{+} + \mbox{H}_{2}  \longrightarrow \mbox{CH}^{+} + \mbox{H},
\end{equation}

\noindent which is endothermic and only occurs at temperatures above $4000$ K~\cite{larsonOrel08,plasil2011,william92,batespitz51,pan2005} or at non thermal collision energies. It is destroyed by reactions with atoms and molecules and by recombination with electrons. There are still many open questions concerning the formation and destruction of CH$^{+}$, as standard gas phase models have not been able to account for the abundance of this molecular ion in diffuse clouds (see e.g.~\cite{godard2013} and references therein). In these environments, the temperature and the pressure are very low and, due to high electron densities, the main route for its destruction is the dissociative recombination (DR) \cite{herbst2005}. This issue has recently been considered in detail by Faure {\it et al}~\cite{faure2017} who showed that the proper treatment of non-local thermodynamic equilibrium effects are essential to model the production of CH$^+$ ions in the ISM.

CH$^{+}$ is the positive ion of the key radical in low temperature low pressure technological plasmas, CH,  used in diamond deposition and responsible for the reactivity of molecular species in the these environments~\cite{hassouni2010}. Understanding its formation and destruction routes can be of great importance in augmenting the deposition rates in these technologies. 

CH$^{+}$ is also a molecular system of interest for the fusion device modelling. As a matter of fact, in the first stage of the operations in the International Thermonuclear Reactor (ITER), the divertor target was made of {\it carbon fibre-reinforced carbon composite} (CFC) \cite{iter2014}.  Since then, the plasma facing material of the reactor was changed to beryllium and tungsten, but in other fusion reactors like JET-C the carbon persist to be a major constituent~\cite{jet}. 
Consequently, CH$^{+}$ is among the hydrocarbon species that will be present in the edge plasma formed, as a result of the bombardment with hydrogenic atoms and ions of the {\it plasma-facing material} (PFM), that is the walls \cite{janev2002,janev2003,swaaij2012}. DR is likely to be the main process for the quenching of CH$^{+}$ ions and of the subsequent release of carbon atoms, playing thus an important role in understanding the carbon re-deposition problem. 

The DR rate coefficient  of CH$^{+}$ is a recurrent subject of controversy between theory and experiment. The goals of this work are: (i) to provide a new attempt to accurately evaluate this rate coefficient and, (ii) to produce DR data relevant for the kinetical modeling of astrochemical and fusion (including ITER) relevance.

Neglecting rotational structure and interactions, 
the DR of CH$^+$ in its ground electronic state and in the vibrational state $v_i^+$  writes as:
\begin{equation}\label{eq:DR}
\mbox{CH}^{+}(v_{i}^{+}) + e^{-}\longrightarrow \mbox{C} + \mbox{H}.
\end{equation}
DR is competed by vibrational transitions (VT):
\begin{equation}
\label{eq:scattering}
\mbox{CH}^{+}(v_{i}^{+}) + e^{-} \longrightarrow  \mbox{CH}^{+} (v_{f}^{+}) + e^{-},
\end{equation}

\noindent
i.e.  excitations (VE, when $v_{f}^{+} > v_{i}^{+}$), elastic scattering (ES, when $v_{f}^{+} = v_{i}^{+}$) and de-excitations (VdE  when $v_{f}^{+} < v_{i}^{+}$), where $v_f^+$ stands for the final vibrational quantum number.

Early calculations of potential energy curves of CH$^+$ were reported by Lorquet {\it et al.} \cite{lor1971}. Potential energy curves of CH$^+$and CH were also produced by Giusti-Suzor and Lefebvre-Brion \cite{gl1977}, but these latter authors did not find a favorable crossing of the CH$^+$ curve by a neutral one. While considering the destruction mechanism of CH$^+$, Solomon and Klemperer \cite{sk1972} (see also Larsson and Orel \cite{larsonOrel08}) assumed an equal dielectronic and dissociative recombination rate coefficient for  CH$^+$. This led to a slow dissociative recombination rate coefficient. However, theoretical calculations by Bardsley and Junker \cite{bj1973} gave a dissociative recombination rate three times higher than the dielectronic recombination rate. The issue was not settled until a very careful calculation was performed by Takagi, Kosugi and Le Dourneuf in 1991 \cite{tkd1991}, who obtained the relevant molecular structure data using the self consistent field (SCF) and the configuration mixing (CM) methods, and were able to identify a neutral $^2\Pi$ state which crosses the ion close to its ground vibrational level and goes to the C($^3$P) + H(1s) asymptotic limit. Starting from these data, and using the multichannel quantum defect  theory (MQDT), they obtained a low-temperature rate constant which was in reasonable agreement with that predicted by Bardsley and Junker \cite{bj1973} and the experimental value given by Mitchell and McGowan~\cite{mm1978}.

On the experimental side, this first measurement of dissociative recombination \cite{mm1978} was followed by a new one of the same team \cite{mulEtal1981}, resulting in slightly larger values. More accurate thermal rate coefficients, close to the theoretical values obtained by Takagi {\it et al} 
\cite{tkd1991}, have been found later by Mitchell \cite{mitchell1990}.  

A very detailed experimental study of CH$^+$ DR, resulting in cross sections, product branching ratios and angular distributions, was given by Amitay {\it et al.} \cite{amitay1996}. Quite surprisingly, unlike previously observed or theoretically predicted narrow resonances in DR cross sections, their measurements showed several unexpectedly broad and prominent resonances. They have been tentatively attributed to the capture of the incident electron into {\it core-excited} Rydberg states, electronically coupled to the initial electronic continuum and to the final dissociative channel.

In order to understand and characterise the broad resonances revealed in the experiment, Carata {\it et al.} \cite{carata2000}  performed new molecular structure data on CH$^+$ and CH, and included in the available  MQDT-DR approach the effect of these core excited states. Apart from the ground state ($X ^1\Sigma^+$) of CH$^+$,  the next two excited states ($a ^3\Pi$, $A ^1\Pi$) of the ion were included in these calculations, together with their $^2\Pi$ respective Rydberg manifolds. Broad resonances in the resulting cross sections similar to those in the experiments of Amitay {\it et al.} were obtained, confirming the claim of capture of the incident electron into core excited Rydberg states. 
However, the absolute magnitude of the experimental results was not reproduced. With a limited number of ionisation channels (corresponding to few among the lowest vibrational states of the ion) and with interactions computed as the first order perturbative solution of the Lippman-Schwinger equation, the theoretical cross section was at best about one order of magnitude lower than the experimental ones at the low end of the range of incident energy they considered ($0.01-4$ eV). 

Whereas the previous work \cite{carata2000} was meant to {\it qualitatively} confirm the role of the capture into core-excited Rydberg states in the shape of the recombination cross section, a newly revealed mechanism at that time, the aim of the present work is to provide a {\it quantitative} complete approach resulting in accurate cross sections, appropriate for a full comparison with the storage-ring data. In this respect, we have included in our calculations all the available discrete vibrational levels of the three lowest electronic states of the molecular ion, and we computed the relevant interactions in the second order of the Lippman-Schwinger equation. This brings a remarkable improvement in the accuracy of the DR cross sections in the energy range considered.

\begin{figure}[t]
\centering
\includegraphics[width=0.95\columnwidth]{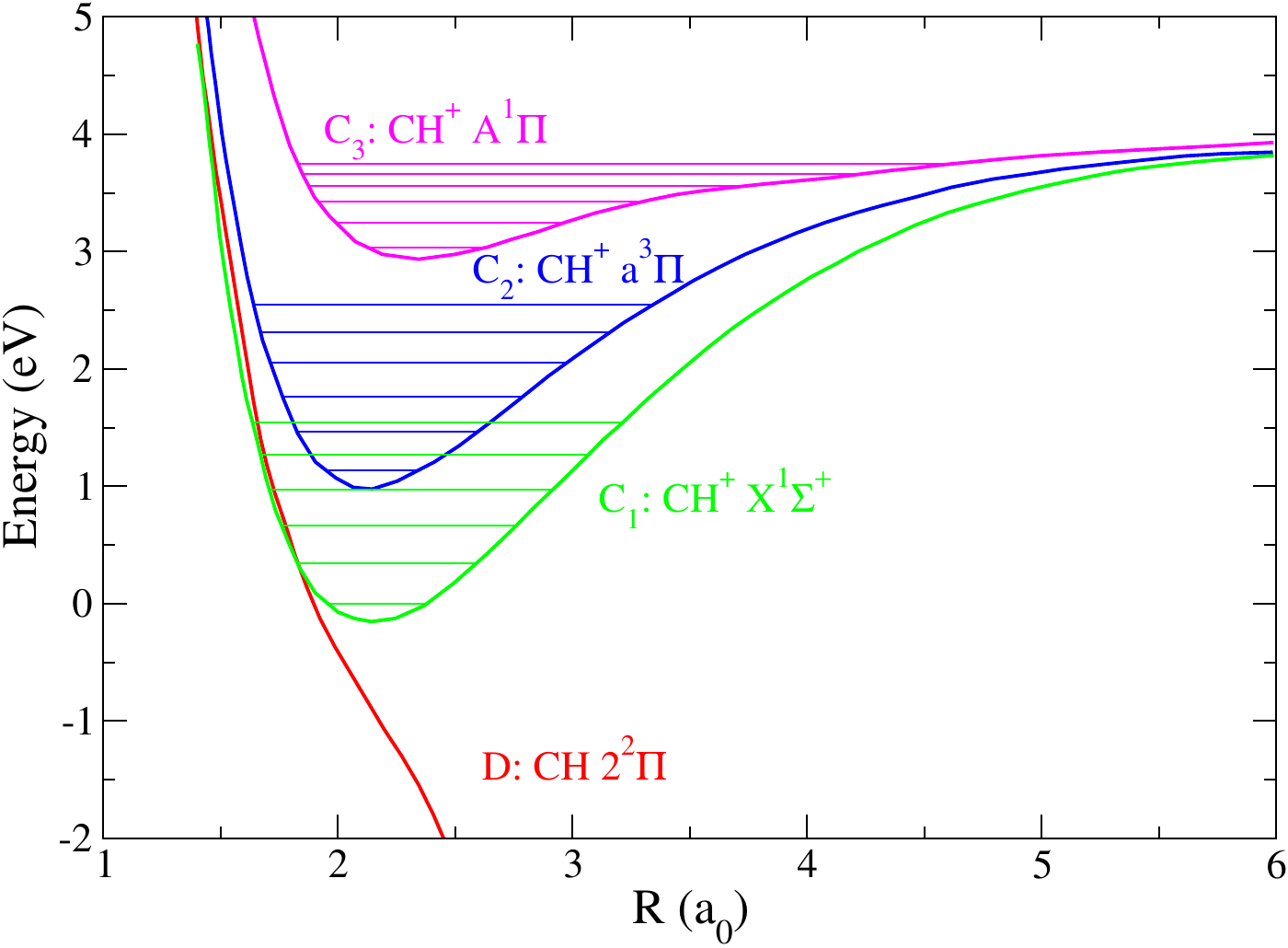}
\caption{
\label{pecs}
(Color online) CH$^+$ and CH states relevant for the electron-CH$^+$ dissociative recombination and resonant vibrational excitation. Shown are the potential energy curves for CH$^+$ ($C_1: X ^{1}\Sigma^{+}$) (lowest curve (green online)), CH$^+$ ($C_2: a ^{3}\Pi$) (middle curve (blue online)), CH$^+$ ($C_3: A ^{1}\Pi$) (topmost curve (magenta online)), and the CH ($D: 2 ^{2}\Pi$)  dissociative state (red online). The lowest vibrational levels of the ground and excited cores are represented by horizontal thin lines, of the same color as those of the respective potential energy curves of the ion cores. The energy scale is relative to the lowest vibrational level of the CH$^+$ ($X ^{1}\Sigma^{+}$) ground state.
}
\end{figure}

\section{Molecular data}

The  minimum set of input data for an MQDT calculation are the potential energy curves
(PEC) of the ground state of the molecular ion, the PECs of the dissociative 
states of the neutral, the electronic couplings of the dissociative states with 
the ionization continuum of the ground state and the quantum defects of the 
Rydberg series converging to the ground state of the ion. 

When further series of core excited Rydberg  states are taken into account, besides their PECs (relying on the quantum 
defects and the PECs of the excited cores), one also needs their mutual (Rydberg-Rydberg) couplings and the couplings of these  series to 
the dissociative states (Rydberg-valence).

All these molecular structure data used in the present calculation are those of Carata {\it et al}~\cite{carata2000}. The PECs of the ion states and of the neutral dissociative ones are shown in Figure \ref{pecs} together with the lowest few vibrational levels of each ionic state. 

\section{The MQDT-type approach to  the low energy dissociative recombination with core excited states}

\subsection{The MQDT formalism for a single ion core}

We restrict ourselves to the case where the energy of the incident electron is lower than the dissociation energy of the target ion, considered to be in its ground electronic state. The collision process involves two mechanisms: 

(a) the direct process, where the incoming electron is captured into a doubly excited neutral dissociative state CH$^{**}$ which either autoionises or leads to C and H neutral fragments:
\begin{equation}\label{eq:direct}
\begin{split}
   \mbox{CH}^{+}(v_{i}^{+})+e^{-}\rightarrow \, & \mbox{CH}^{**}
   \rightarrow 
   \left\{ 
      \begin{aligned}
       &\mbox{C} + \mbox{H}\\
       & \mbox{CH}^{+} (v_{f}^{+}) +e^{-},
     \end{aligned}
  \right.
\end{split}
\end{equation}

(b) the indirect process, where the incident electron is temporarily captured into a singly excited bound Rydberg state CH$^*$, which in turn will be predissociated by CH$^{**}$:
\begin{equation}\label{eq:indirect}
\begin{split}
   \mbox{CH}^{+}(v_{i}^{+})+e^{-}\rightarrow \, &\mbox{CH}^{*}\rightarrow \, \mbox{CH}^{**}
   \\
	\mbox{CH}^{**}\rightarrow & 
   \left\{ 
      \begin{aligned}
       &\mbox{C} + \mbox{H}\\
       & \mbox{CH}^{+} (v_{f}^{+}) +e^{-}.
     \end{aligned}
  \right.
\end{split}
\end{equation}

These processes involve  {\it ionisation} and {\it dissociation} channels, {\it closed} if the total energy of the molecular system is lower than the energy of its fragmentation threshold, and  {\it open} in the opposite case. The closed channels are responsible for the indirect mechanism via a series of Rydberg states, differing only by the principal quantum number of the external electron. The interference between this mechanism and the direct one results in the total process \cite{ifs94}.

The dissociation channels $d_j$ are coupled to the ionisation channels relying on the ground-state electronic core ($c_1$) by the short range Rydberg-valence couplings, expressed by \cite{giusti1980}:
\begin{equation}\label{eq:elcoup}
\MV^{(e)\Lambda}_{d_{j},c_1}(R) = \langle\Phi_{d_j}|H_{el}|\Phi^{el,c_1}\rangle,
\end{equation}

\noindent where $\MV^{(e)\Lambda}_{d_{j},c_1}(R)$ is assumed to be independent of the energy of the external electron and the integration is performed over the electronic coordinates of the neutral (electron plus ion({\it core})) system. Here $H_{el}$ denotes the electronic Hamiltonian, $\Lambda$ the absolute value of the projection of the electronic angular momentum of the neutral system on the internuclear axes, $\Phi_{d_j}$ is the electronic wave function of the dissociative state $d_j$, and $\Phi^{el,c_1}$ the wave function describing the electron-ion system, and it is assumed that one single partial wave of the incident electron contributes to this interaction.

Integrating these couplings over the internuclear distance, the non-vanishing elements of the interaction matrix $\V(E)$ are:
\begin{equation}\label{eq:Vdv}
V_{d_j,v_{c_1}}^{\Lambda}(E) = \langle F_{d_j}(E)|\MV^{(e)\Lambda}_{d_{j},c_1}(R)|\chi_{v_{c_1}}
	\rangle.
\end{equation}

\noindent
Here $\chi_{v_{c_1}}$ is the vibrational wave function associated with an ionisation channel in the reaction zone, $F_{d_j}$ is the regular radial wave function of the dissociative state $d_j$ and $E$ is the total energy of the molecular system. This interaction is effective at short electron-ion and internuclear distances typical of the reaction zone.

Starting with the interaction matrix $\Vmat$, a short range reaction matrix $\Kmat$ is then built as a solution of the Lippmann-Schwinger equation
\begin{equation}\label{eq:LS}
\boldsymbol{{\MK}}= \boldsymbol{\Vmat} + \boldsymbol{\Vmat} \frac{1}{E-\boldsymbol{H_0}} \boldsymbol{\MK},
\end{equation}

\noindent where $\boldsymbol{H_0}$ is the zero-order Hamiltonian of the molecular system. 

The structure of the reaction matrix $\Kmat$ in block form is the following:
\begin{equation}\label{Kmat_1core}
\Kmat = \left( \begin{array}{cc}
\Kmat_{\bar d\bar d} & \Kmat_{{\bar d}\bar v_{c_1}} \\
\Kmat_{\bar v_{c_1} \bar d} & \Kmat_{\bar v_{c_1} \bar v_{c_1}} \\
\end{array} \right),
\end{equation}

\noindent where the collective indices $\bar d$ and $\bar v_{c_1}$ span the ensembles of all individual indices $d_j$ and $v_{c_1}$ which respectively label dissociation channels and ionisation channels built (the latter ones) on core $C_1$. 

The second order solution of Eq. (\ref{eq:LS}) is exact, providing that we neglect the energy dependence  of the coupling matrix $\Vmat(R)$ \cite{ngassam2003b}:
\begin{equation}\label{Kmat_core1}
\Kmat = \left( \begin{array}{cc}
\Omat & \V_{\bar d \bar v_{c_1}} \\
\V_{\bar v_{c_1} \bar d} & \Kmat^{(2)}_{\bar v_{c_1} \bar v_{c_1}} \\
\end{array} \right).
\end{equation}  

\noindent where the elements of the diagonal block $\Kmat^{(2)}_{\bar v_{c_1} \bar v_{c_1}}$ of $\Kmat$ are  \cite{ifs91}:
\begin{eqnarray}
\Kmat^{(2)}_{v_{c_1} v'_{c_1}} = && \sum_{d_j}\frac{1}{W_{d_j}} \int \int \Big[\chi_{v_{c_1}}^\Lambda(R)
\MV_{d_j,c_1}^{(e)\Lambda} (R) F_{d_j} (R_<)\times\nonumber \\
&\times&G_{d_j}(R_>) \MV_{d_j,c_1}^{(e)\Lambda} (R') \chi_{v'_{c_1}}^\Lambda 
(R') \Big]dR dR',
\end{eqnarray}
\noindent $\Omat$ is the null matrix, and $W_{d_j}$ is the Wronskian of the regular and irregular pair ($F_{d_j}$, $G_{d_j}$).

To express the effects of short range interactions in terms of phase shifts, we diagonalise the $\Kmat$ matrix
\begin{equation}\label{K-pvp}
\Kmat \mbox{\boldmath $\mathcal{U}$} = -\frac{1}{ \pi}\; 
\boldsymbol{\tan(\mbox{$\eta$})} \mbox{\boldmath $\mathcal{U}$},
\end{equation}

\noindent where $\Umat$ is a matrix whose columns are eigenvectors of matrix $\Kmat$ and the diagonal matrix $\tan(\etamat)$ contains its eigenvalues.

In the external region \cite{jungenAtabek1977}, where the Born-Oppenheimer representation is no longer valid for the neutral molecule, a frame transformation \cite{changFano1972,giusti1980} is performed via the projection coefficients
\begin{equation}\label{C1}
\mathcal{C}_{v^+_{c_1},\Lambda \alpha} = \sum_{v_{c_1}} U^{\Lambda}_{v_{c_1}, \alpha}\langle 
\chi_{v^{+}_{c_1}} (R)| \cos(\pi\mu^{\Lambda}_{c_1} (R) + \eta_{\alpha}^{{\Lambda}})|
\chi_{v_{c_1}}(R) \rangle
\end{equation}
\begin{equation}\label{C2}
\mathcal{C}_{{d_j},\Lambda\alpha} = U^{\Lambda}_{{d_j}, \alpha} \cos 
\eta^{\Lambda}_\alpha
\end{equation}
\begin{equation}\label{S1}
\mathcal{S}_{v^+_{c_1},\Lambda \alpha} = \sum_{v_{c_1}} U^{\Lambda}_{v_{c_1}, \alpha}\langle 
\chi_{v^{+}_{c_1}} (R)| \sin(\pi\mu^{\Lambda}_{c_1} (R) + \eta_{\alpha}^{{\Lambda}})|
\chi_{v_{c_1}}(R) \rangle
\end{equation}
\begin{equation}\label{S2}
\mathcal{S}_{{d_j},\Lambda\alpha} = U^{\Lambda}_{{d_j}, \alpha} \sin 
\eta^{\Lambda}_\alpha,
\end{equation}

\noindent where $\alpha$ denotes the eigenchannels built through the diagonalization of the reaction matrix $\Kmat$. These can be grouped into the matrices $\Cmat$ and $\Smat$, which are the building blocks of the generalized scattering matrix $\Xmat$ that involves all open ("\textit{o}") and closed ("\textit{c}") channels. The $\Xmat$ matrix in turn can be arranged into four sub matrices
\begin{equation}\label{eq:Xmatrix}
\boldmath{
\Xmat =\frac{\Cmat+i\Smat}{\Cmat-i\Smat}=
\left(
  \begin{array}{cc}
    \Xmat_{oo} & \Xmat_{oc} \\
    \Xmat_{co} & \Xmat_{cc} \\
  \end{array}
\right)}.
\end{equation}

Imposing boundary conditions leads to the physical scattering matrix \cite{seaton1983}:
\begin{equation}\label{eq:smatrix}
\mbox{\boldmath$
\SSmat=\Xmat_{oo}-\Xmat_{oc}$}\frac{\mathbf{1}}{\mbox{\boldmath$\Xmat_{cc}$}
-\exp({\rm -i 2 \pi} \mbox{\boldmath$\numat$)}} \mbox{\boldmath$\Xmat_{co}$},
\end{equation}

\noindent where the diagonal matrix $ \numat$ is constructed with the effective quantum numbers $\nu_{v^+_{c_1}} = [2(E_{v^+_{c_1}} - E)]^{-1/2}$ (in atomic units) associated with each vibrational threshold $E_{\nu^+}$ of the ion, situated above the current total energy $E$, labelling a closed channel.

For a molecular ion, initially in the vibrational state ${v^+_{c_1i}}$, recombining with an electron of energy $\varepsilon$, the total cross section of capture into all dissociative states $d_j$, summed over all symmetries ($"sym"$: gerade or ungerade, singlet or triplet) and all available electronic angular momentum projection $\Lambda$ values, we get:
\begin{equation}\label{drxsec1}
	\sigma_{diss \leftarrow {v^+_{c_1i}}}^{DR} = \frac{\pi}{4\varepsilon} \sum_{sym,\Lambda} 
	\rho^{sym, \Lambda} \sum_{j} \left|S_{d_j, {v^+_{c_1i}}}\right|^2.
\end{equation}

\noindent
Here $\rho^{sym, \Lambda}$ is the ratio between the spin multiplicities of the neutral and the target ion.

Similarly, the total cross section for vibrational excitation from the initial vibrational state ${v^+_{c_1i}}$ into the state ${v^+_{c_1f}}$ is given by:
\begin{equation}\label{vexsec1}
	\sigma_{{v^+_{c_1f}} \leftarrow  {v^+_{c_1i}}}^{VE} = \frac{\pi}{4\varepsilon}\sum_{sym,\Lambda}
\rho^{sym, \Lambda} \left|S_{{v^+_{c_1f}}, {v^+_{c_1i}}} - \delta_{{v^+_{c_1f}}{v^+_{c_1i}}}\right|^2.
\end{equation}

\subsection{Inclusion of core excited states}

The MQDT formalism in the previous section is valid for a system where one or more dissociative states are coupled to the ionisation channels of the ground ion core. However, CH$^+$ has several bound excited states whose ionisation continua are coupled to the ionisation continuum of the ground core and to  neutral dissociative states. 

For the energy range characterising the incident electron in the present work, two such excited states are relevant, i.e. those of $a\,^{3}\Pi$ and $A\,^{1}\Pi$ symmetry, which we respectively call core 2 and core 3, labelled $C_2$ and $C_3$ in fig.~\ref{pecs}. The neutral $2\, ^{2}\Pi$  dissociative state is coupled to the ionisation channels of the three ion cores and is mainly responsible for driving the low energy DR mechanism. 

The interaction between the ionisation and dissociation channels result in two types of couplings, namely the Rydberg-valence couplings 
\begin{equation}\label{eq:Vdv_c2}
V_{d_j,v^+_{c_2}}^{\Lambda}(E) = \langle F_{d_j}(E)|\MV^{(e)\Lambda}_{d_{j},c_2}(R)|\chi_{v^+_{c_2}}
	\rangle,
\end{equation}
\begin{equation}\label{eq:Vdv_c3}
V_{d_j,v^+_{c_3}}^{\Lambda}(E) = \langle F_{d_j}(E)|\MV^{(e)\Lambda}_{d_{j},c_3}(R)|\chi_{v^+_{c_3}}
	\rangle,
\end{equation}

\noindent and the Rydberg-Rydberg couplings given by 
\begin{equation}\label{eq:Vvw_c1c2}
V^{\Lambda}_{v^+_{c_1}v^+_{c_2}} = \langle \chi_{v^+_{c_1}} |\widetilde{\MV}^{(e)\Lambda}_{c_{1},c_2}(R)|\chi_{v^+_{c_2}} \rangle,
\end{equation}
\begin{equation}\label{eq:Vvw_c1c3}
V^{\Lambda}_{v^+_{c_1}v^+_{c_3}} = \langle \chi_{v^+_{c_1}} |\widetilde{\MV}^{(e)\Lambda}_{c_{1},c_3}(R)|\chi_{v^+_{c_3}} \rangle,
\end{equation}
\begin{equation}\label{eq:Vvw_c2c3}
V^{\Lambda}_{v^+_{c_2}v^+_{c_3}} = \langle \chi_{v^+_{c_2}} |\widetilde{\MV}^{(e)\Lambda}_{c_{2},c_3}(R)|\chi_{v^+_{c_3}} \rangle,
\end{equation}

\noindent where the different electronic couplings are defined by eq.~(\ref{eq:elcoup}) with the logical interchange of the ion cores, while the vibrational quantum numbers $v^+_{c_i}$, label the ionisation channels associated to the core $i$, with $i=1,2,3$.

The structure of the reaction matrix $\Kmat$ in block form is the following:
\begin{equation}\label{Kmat2}
\Kmat = \left( \begin{array}{cccc}
\Kmat_{\bar d\bar d} & \Kmat_{{\bar d}\bar v^+_{c_1}} & \Kmat_{\bar d \bar v^+_{c_2}}
& \Kmat_{\bar d \bar v^+_{c_3}}\\
\Kmat_{\bar v^+_{c_1} \bar d} & \Kmat_{\bar v^+_{c_1} \bar v^+_{c_1}} & \Kmat_{\bar v^+_{c_1} \bar v^+_{c_2}}
& \Kmat_{\bar v^+_{c_1} \bar v^+_{c_3}}\\
\Kmat_{\bar v^+_{c_2} \bar d} & \Kmat_{\bar v^+_{c_2} \bar v^+_{c_1}} & \Kmat_{\bar v^+_{c_2} \bar v^+_{c_2}}
& \Kmat_{\bar v^+_{c_2} \bar v^+_{c_3}}\\
\Kmat_{\bar v^+_{c_3} \bar d} & \Kmat_{\bar v^+_{c_3} \bar v^+_{c_1}} & \Kmat_{\bar v^+_{c_3} \bar v^+_{c_2}}
& \Kmat_{\bar v^+_{c_3} \bar v^+_{c_3}}\\
\end{array} \right),
\end{equation}

\noindent where the collective indices $\bar d$, $\bar v^+_{c_i}$, $i=1,2,3$ span the ensembles of all individual indices connected to the dissociation channels and ionisation channels built on $C_1$, $C_2$ and $C_3$ ion cores. 

\begin{figure}[t]
\begin{center}
\includegraphics[width=0.95\columnwidth]{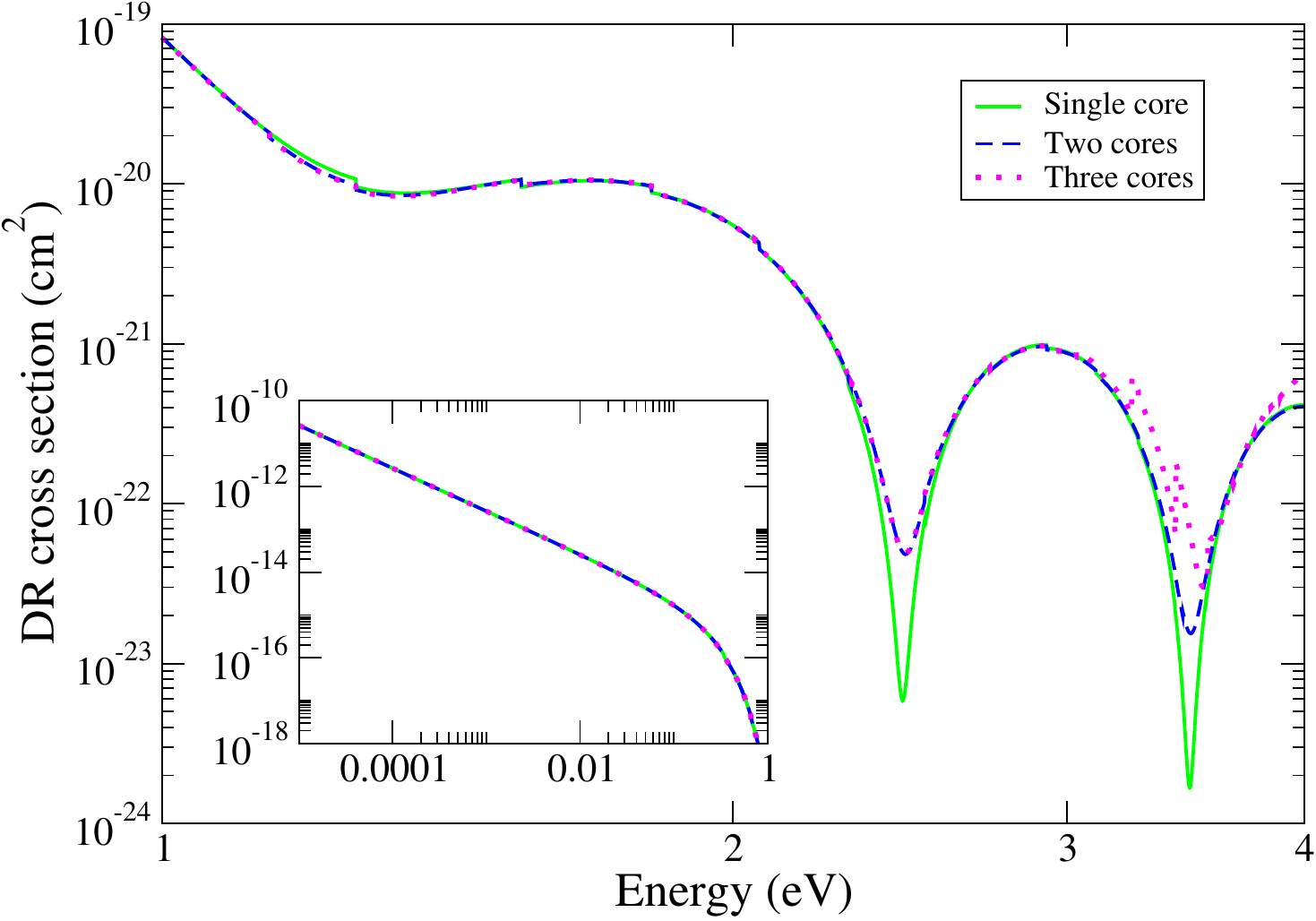}
\caption{
\label{dir}
(Color online) First order direct dissociative recombination of CH$^+$ in its ground state $X^2\Sigma^+$ ($v_i^+ = 0$) for three cases involving the consecutive inclusion of the ground (in green), first excited $C_2$ (in blue) and second excited $C_3$ (in magenta) state of the ion core
represented in Fig.~\ref{pecs}. The figure shows the cross sections in the higher energy range, whereas the inset gives them in the lower one. 
}
\end{center}
\end{figure}

An extensive and rigorous derivation of the structure of each block of the $\Kmat$-matrix in second order was first provided in our earlier study \cite{kc2013a} on H$_2^+$ and HD$^+$ with two cores, where in addition to the H$_2^+$ and HD$^+$ ground core, a repulsive ion core was present and contributed to the process. This was further developed for the N$_2^+$~\cite{little2014} molecular system with three attractive ion cores. More recently it was applied for the SH$^+$~\cite{mezei2017} molecular cation, which shows very similar formation pathways to CH$^+$ in the interstellar environment. The natural extensions of our earlier works lead to the following form of the $\Kmat$-matrix in second order,
\begin{equation}\label{Kmat3}
\Kmat = \left( \begin{array}{cccc}
\Omat & \V_{\bar d \bar v^+_{c_1}} & \V_{\bar d \bar v^+_{c_2}} & \V_{\bar d \bar v^+_{c_3}}\\
\V_{\bar v^+_{c_1} \bar d} & \Kmat_{\bar v^+_{c_1} \bar v^+_{c_1}}^{(2)} & \V_{\bar v^+_{c_1} \bar v^+_{c_2}}
& \V_{\bar v^+_{c_1} \bar v^+_{c_3}}\\
\V_{\bar v^+_{c_2} \bar d} & \V_{\bar v^+_{c_2} \bar v^+_{c_1}} & \Kmat_{\bar v^+_{c_2} \bar v^+_{c_2}}^{(2)}  
& \V_{\bar v^+_{c_2} \bar v^+_{c_3}}\\
\V_{\bar v^+_{c_3} \bar d} & \V_{\bar v^+_{c_3} \bar v^+_{c_1}} & \V_{\bar v^+_{c_3} \bar v^+_{c_2}}  
& \Kmat_{\bar v^+_{c_3} \bar v^+_{c_3}}^{(2)}\\
\end{array} \right),
\end{equation}  

\noindent where the elements of the diagonal blocks of the $\Kmat$ write:
\begin{eqnarray}
K_{ii'}^{(2)} = &&\sum_{d_j} \frac{1}{W_{d_j}} \int \int \Big[\chi_{i}^\Lambda(R)
\MV_{d_j} (R) F_{d_j} (R_<)\times\nonumber \\
&\times&G_{d_j}(R_>) \MV_{d_j} (R') \chi_{i'}^\Lambda 
(R') \Big]dR dR',
\end{eqnarray}

\noindent 
$i$ and $i'$ standing for vibrational quantum numbers $v^+_{c_i}$, with $i=1,2,3$, and $ \MV_{d_j}(R)$ is a simplified notation for the couplings appearing in eqs.~(\ref{eq:elcoup}),~(\ref{eq:Vdv_c2}) and~(\ref{eq:Vdv_c3}).

\subsection{Calculation of cross sections}

The inclusion of the additional excited ion cores increases not only the dimension of the $\Kmat$-matrix~(Eq. (\ref{Kmat_core1})), but also that of the $\Cmatt$ and $\Smatt$ matrices. More specifically, besides the matrix elements given by Eqs.~(\ref{C1}) and~(\ref{S1}), further matrix elements related to cores 2 and 3 contribute to the building of these matrices:
\begin{equation}\label{C1_core2}
\mathcal{C}_{v^+_{c_2},\Lambda \alpha} = \sum_{v_{c_2}} U^{\Lambda}_{v_{c_2}, \alpha}\langle 
\chi_{v^+_{c_2}} (R)| \cos(\pi\mu^{\Lambda}_{c_2} (R) + \eta_{\alpha}^{{\Lambda}})|
\chi_{v_{c_2}}(R) \rangle
\end{equation}
\begin{equation}\label{S1_core2}
\mathcal{S}_{v^+_{c_2},\Lambda \alpha} = \sum_{v_{c_2}} U^{\Lambda}_{v_{c_2}, \alpha}\langle 
\chi_{v^+_{c_2}} (R)| \sin(\pi\mu^{\Lambda}_{c_2} (R) + \eta_{\alpha}^{{\Lambda}})|
\chi_{v_{c_2}}(R) \rangle
\end{equation}
\begin{equation}\label{C1_core3}
\mathcal{C}_{v^+_{c_3},\Lambda \alpha} = \sum_{v_{c_3}} U^{\Lambda}_{v_{c_3}, \alpha}\langle 
\chi_{v^+_{c_3}} (R)| \cos(\pi\mu^{\Lambda}_{c_3} (R) + \eta_{\alpha}^{{\Lambda}})|
\chi_{v_{c_3}}(R) \rangle
\end{equation}
\begin{equation}\label{S1_core3}
\mathcal{S}_{v^+_{c_3},\Lambda \alpha} = \sum_{v_{c_3}} U^{\Lambda}_{v_{c_3}, \alpha}\langle 
\chi_{v^+_{c_3}} (R)| \sin(\pi\mu^{\Lambda}_{c_3} (R) + \eta_{\alpha}^{{\Lambda}})|
\chi_{v_{c_3}}(R) \rangle
\end{equation}

\noindent and, consequently, via Eqs.~(\ref{eq:Xmatrix}) and~(\ref{eq:smatrix}), to the building of the matrices \mbox{\boldmath{$\Xmat$}} and \mbox{\boldmath{$\SSmat$}}. The DR and the vibrational excitation cross sections are then obtained from equations~(\ref{drxsec1}) and~(\ref{vexsec1}).

Figure~\ref{dir} shows the first order direct DR cross sections of CH$^+$ in its ground state $X ^2\Sigma^+$ $v_i^+ = 0$, on a wide range of collision energy. One can observe that there is no significant contribution to this cross section from core 2 and core 3 up to approximately $2.3$ eV electron energy, and that we get visible differences between the cross sections whenever the vibrational levels of the excited cores become energetically accessible.

\begin{figure}[t]
\begin{center}
\includegraphics[width=0.95\columnwidth]{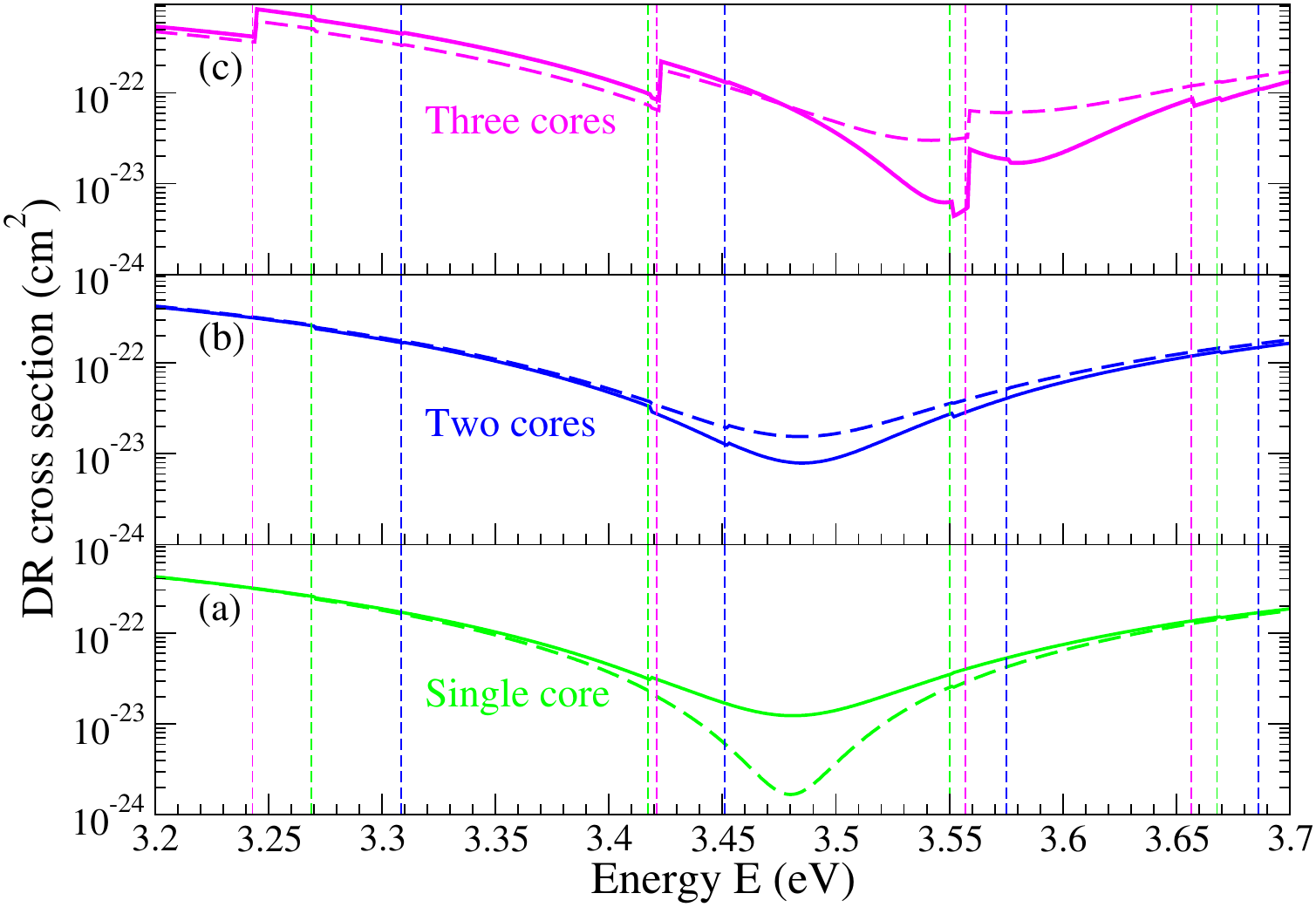}
\caption{
\label{dir12ord}
(Color online) Enlarged view of first order (dashed) and second order (continuous) direct DR cross sections of CH$^+$ in its ground state $X ^2\Sigma^+$ ($v_i^+ = 0$), with the number of cores used in the calculations shown in each panel. The energy range chosen is limited to the region where the first and second order cross sections show the most significant difference. The vibrational levels of $C_1$, $C_2$ and $C_3$ are shown as vertical dashed lines using the same color code as that of  Fig.~\ref{pecs}.
}
\end{center}
\end{figure}

We can see in the enlarged view presented in figure~\ref{dir12ord} the importance of the second order treatment of the reaction matrix . The energy range of the figure was chosen such that to obtain the most significant differences for the consecutive addition of the excited cores into the calculation.

\section{Results and discussion}

In this section we present our results for the DR cross sections, rate coefficients and vibrational excitations from the lowest vibrational level.

The present calculation includes a total of $42$ ionisation channels associated with $19$ vibrational levels of the CH$^+$ ($X ^{1}\Sigma^{+}$) ground state and $14$ and $9$ vibrational levels of CH$^+$ ($A ^{3}\Pi$) (core 2) and CH$^+$ ($a ^{1}\Pi$) (core 3) respectively.

The incident electron energy range is $0.01$ - $0.5$ eV, which is typical for the interstellar environments. Moreover, a common temperature of around $3000 K$ in the divertor region of the present fusion devices indicates that the chosen energy range is also suitable for studying DR in the edge cold plasma.

\subsection{DR cross sections}

\begin{figure}[t]
\begin{center}
\includegraphics[width=0.95\columnwidth]{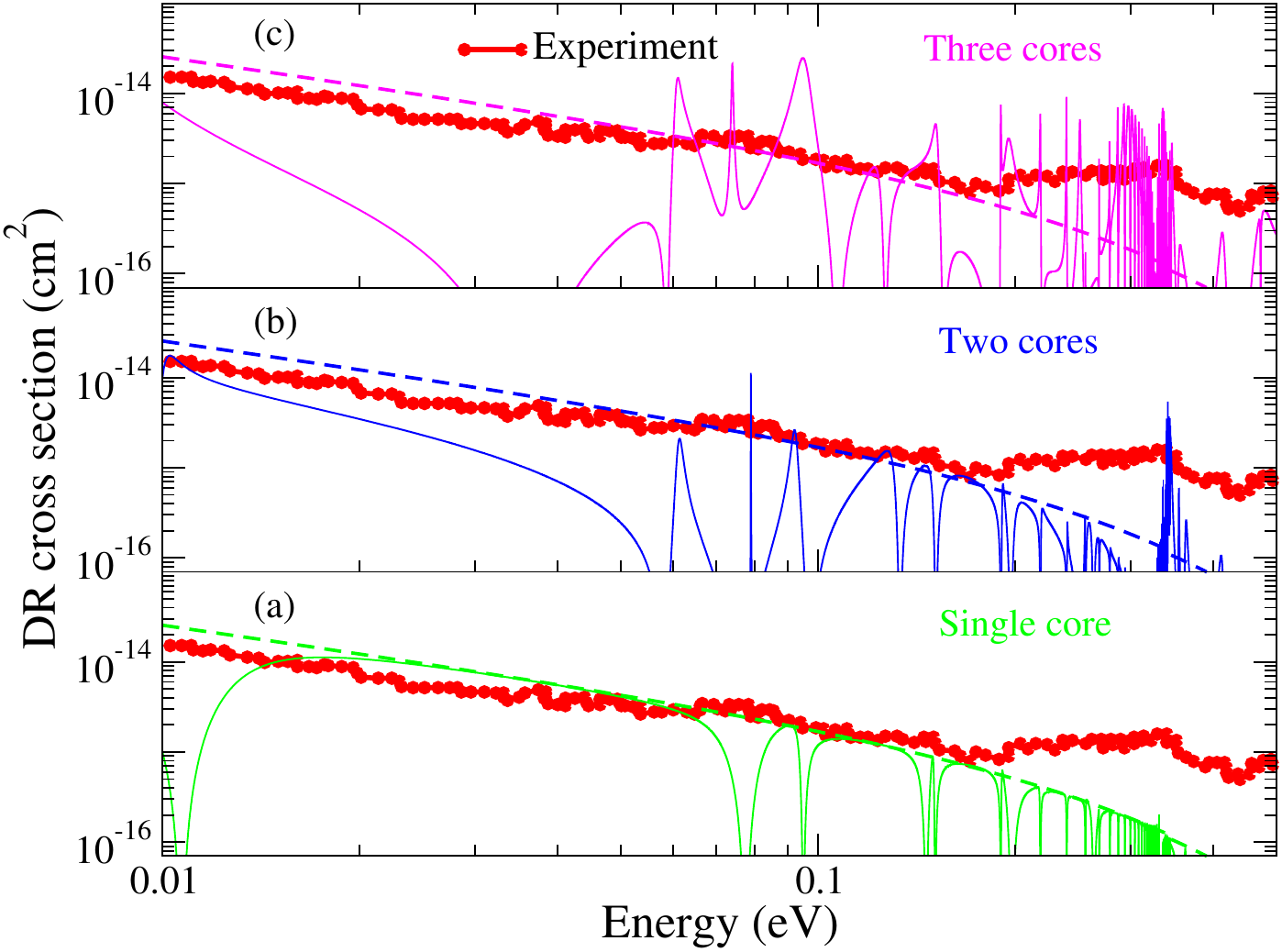}
\caption{
\label{c123tot}
(Color online) First order direct (dashed) and total (continuous) DR cross sections of CH$^+$ ground state $X ^2\Sigma^+$ ($v_i^+ = 0$) (green: single core; blue: two cores; magenta: three cores). Experiment from Amitay \textit{et al.} \cite{amitay1996}.
}
\end{center}
\end{figure}

The DR cross section changes substantially from that of Figs.~\ref{dir} and \ref{dir12ord} 
when we consider the effect of {\it closed} channels - {\it indirect} process - included in the {\it total} cross sections, and progressively take into account the effect of the core excited states. The resulting total cross section is governed by resonance structures, due to the temporary capture into bound highly excited Rydberg states correlating to the ground and to the two lowest excited electronic states of the ion $C_1,C_2$ and $C_3$.

In Figure \ref{c123tot} the total cross section for a single core is shown in the bottom panel (a). One may notice the huge gap between the theoretical result and the experimental one above 200 meV. In the panel (b) we include the contribution of the second core and panel (c) shows the effect of including all the three CH$^+$ ion cores on the DR cross sections. Unlike first order direct cross sections, the first order total cross sections show significant changes as each ion core is included in the calculation and the gap between theory and experiment is progressively (but still partially) filled. When all the three cores are included, the results above 70 meV are clearly closer to the experiments.

\begin{figure}[t]
\begin{center}
\includegraphics[width=0.9\columnwidth]{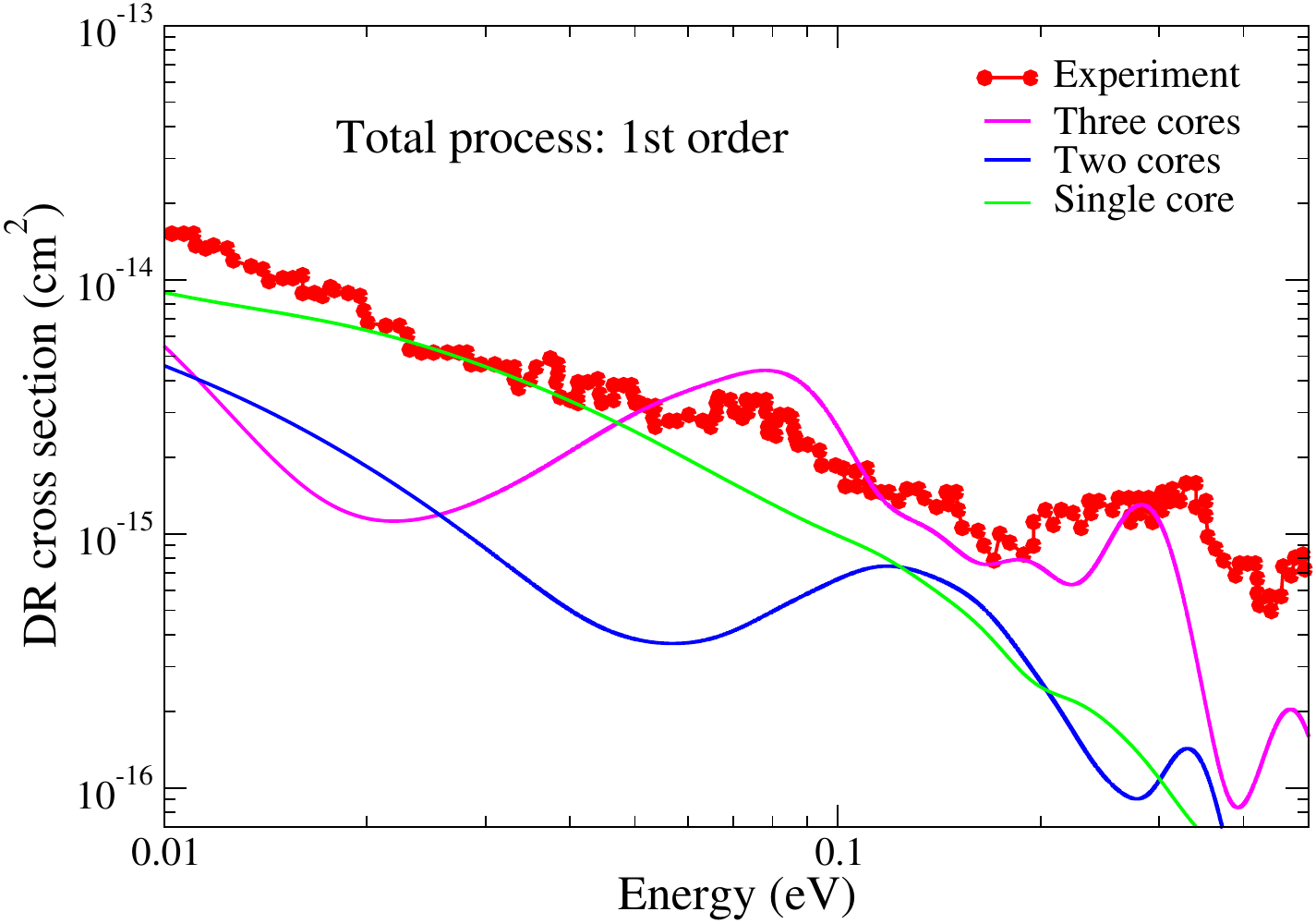}
\caption{
\label{c123conv}
(Color online) First order CH$^+$  DR cross sections after convolution with anisotropic Maxwell distribution. Continuous curves, convoluted cross sections (green: single core; blue: two cores; magenta: three cores). The longitudinal temperature is $kT_\parallel = 1$ meV and the transversal temperature $kT_\perp = 17$ meV. The experiment is from Amitay  \textit{et al.} \cite{amitay1996}.
}
\end{center}
\end{figure}

In order to have a clearer comparison with the experiments we have convoluted our calculated total cross sections with the anisotropic Maxwellian distribution given by 
\begin{equation}\label{Maxwell}
f(v_d, \vec{v}) = \frac{m}{2 \pi k T_{\perp}}\sqrt{\frac{m}{2 \pi k T_\perp}}
exp\Big[-\frac{mv^2_\perp}{2kT_\perp} -\frac{m(v_\parallel - v_d)^2}
{2kT_\parallel} \Big],
\end{equation}

\noindent where $m$ is the electron mass, $v_d$ is the average velocity of the electrons with respect to that of the ions, called the detuning velocity, $v_\perp$ and $v_\parallel$ are respectively the transversal and longitudinal components of the relative electron-ion velocity $\vec{v}$ and $T_\perp$ and $T_\parallel$ are the temperatures associated with the relative motion in the transversal and longitudinal degrees of freedom. The obtained anisotropic rates have been, in accordance with \cite{amitay1996}, divided by the velocity of the electrons, leading to DR cross sections. The convolution procedure smooths out the numerous narrow resonances, typical of the capture of the incoming electron into high $n$ neutral Rydberg states associated with the ground core C$_1$, and leaves behind broad resonances that occur due to the presence of core excited states  C$_2$ and  C$_3$.

Figure \ref{c123conv} shows the first order results after convolution, when one, two and three cores are included progressively. The results with three cores are clearly close to the experimental ones between 50 et 300 meV, and reproduce well the measured broad structures.

\begin{figure}[t]
\begin{center}
\includegraphics[width=0.95\columnwidth]{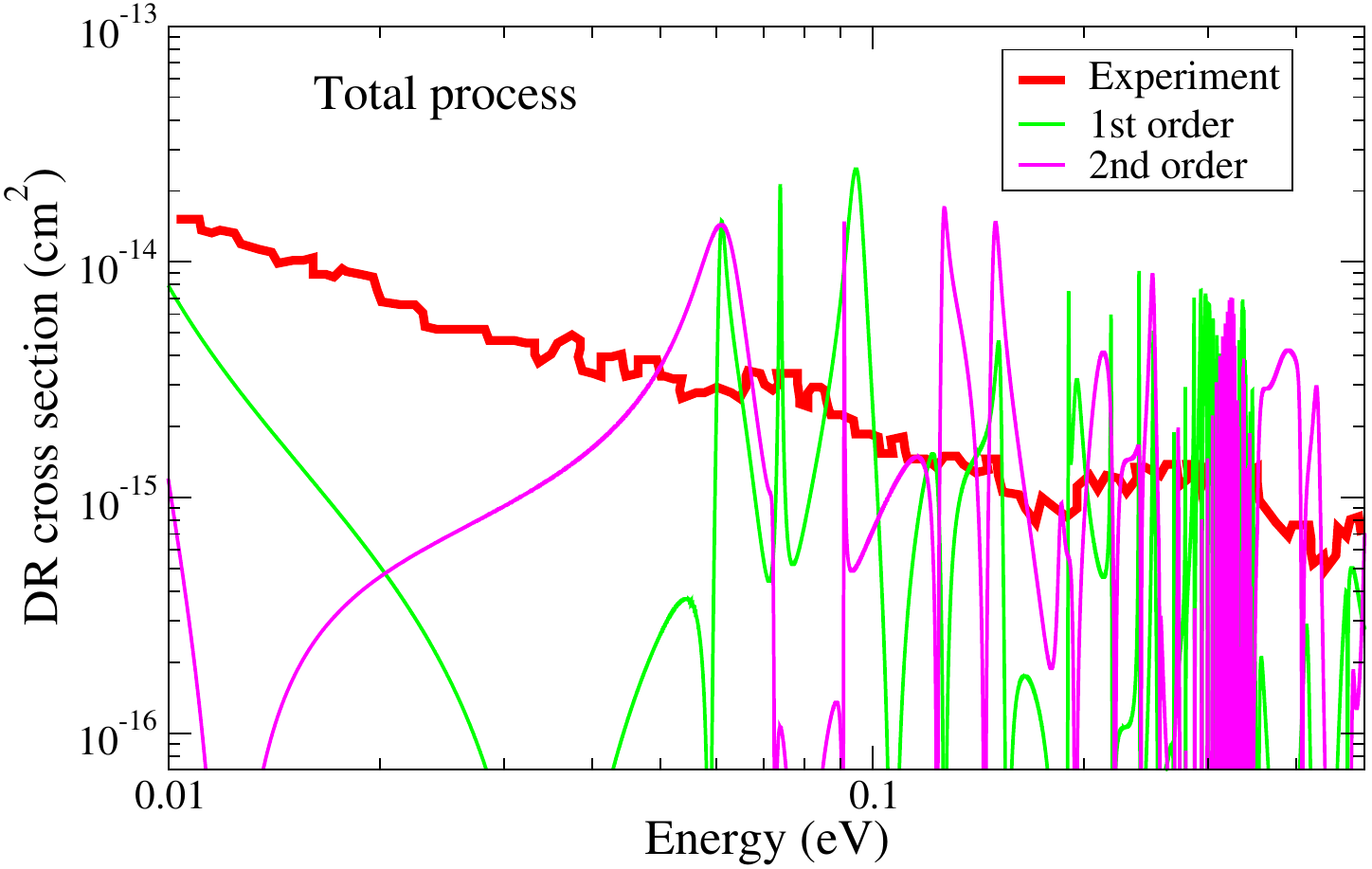}
\caption{
\label{totc3}
(Color online) First order (green) and second order (magenta) total DR cross sections of CH$^+$ in its ground state X$^2\Sigma^+$ ($v_i^+ = 0$) calculated with 3 cores.
}
\end{center}
\end{figure}

\begin{figure}[t]
\begin{center}
\includegraphics[width=0.95\columnwidth]{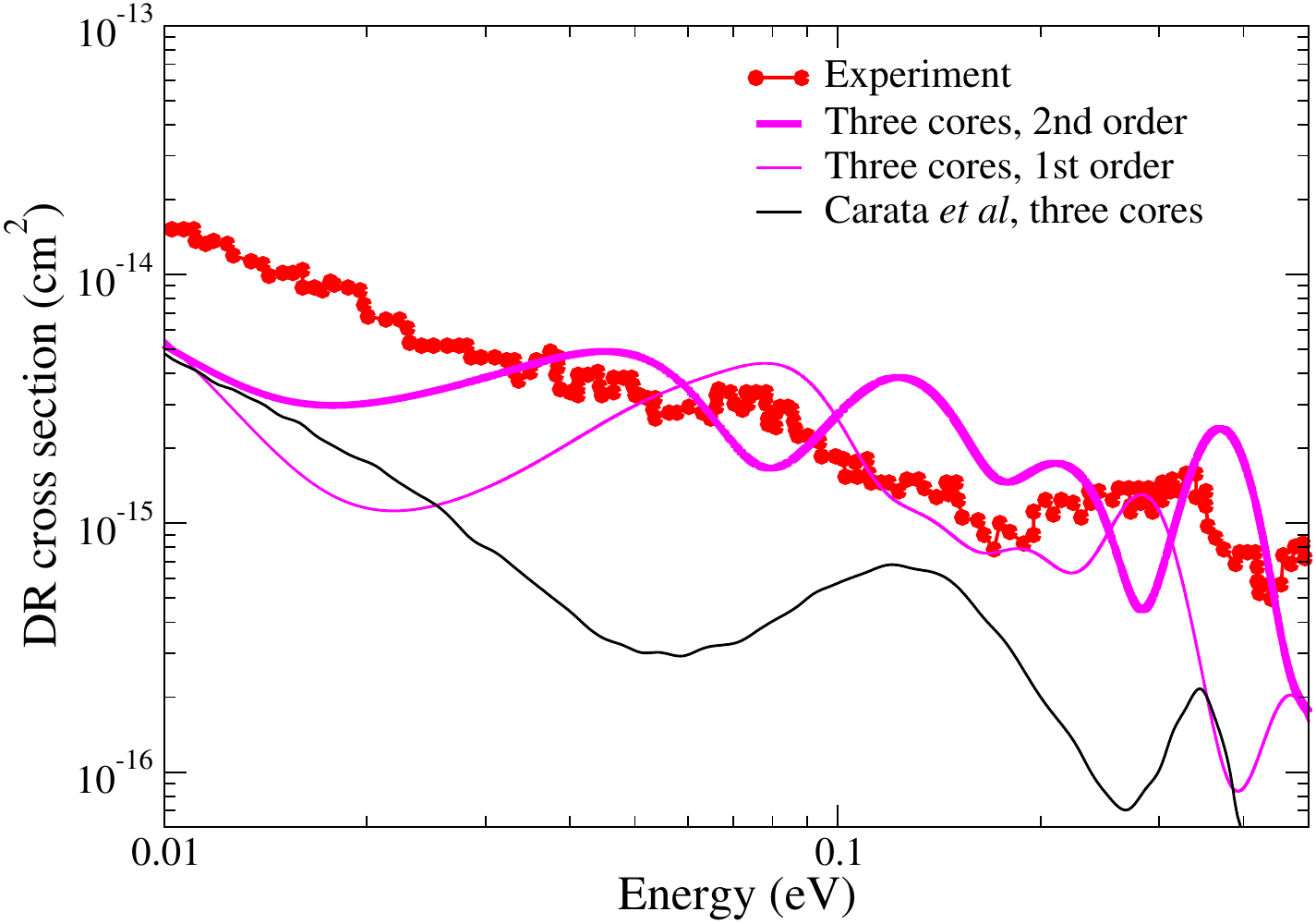}
\caption{
\label{c123conv2}
(Color online) CH$^+$  DR cross sections after convolution with anisotropic Maxwell distribution. Thin curves, first order results with three cores. Thick curve, second order results with three cores. The lowest curve (thin black online) represents the results of Carata {\it et al.} \cite{carata2000}. The experiment is from Amitay \textit{et al.} \cite{amitay1996}.
}
\end{center}
\end{figure}

In Figure \ref{totc3}, the first and the second order total cross sections are presented for the case of  three ion cores included.
We may notice that there is a significant difference between the two, though a quantitative comparison is not possible due to rapid oscillation in the cross section. The comparison between our first and second order {\it convoluted} results including all the three cores with the experiments \cite{amitay1996} are shown in Figure \ref{c123conv2}, together with the first order results of Carata {\it et al.} \cite{carata2000}. These latter results were calculated in the first order of the $K$-matrix with the closed channels corresponding to the lowest vibrational states of the ion only included for each of the ion cores, whereas the present MQDT calculation accounts all the closed channels related to the three core states of the ion and evaluates the $K$-matrix in the second order.

 Performing the calculations in the second order, the cross section is clearly improved over our first order ones, and agree much better with the experiments, especially at higher collision energies. The first order results of Carata {\it et al.} are much lower than our first order results. 

One can conclude that the inclusion of all the closed channels (indirect process) and of the excited cores clearly have improved the agreement with the experimental results, and provided better theoretical cross sections than the existing ones over a wide collision energy range above $30$ meV. Below $30$ meV, the calculation underestimates the experimental cross section up to a factor of $3$.

\begin{figure}[t]
\begin{center}
\includegraphics[width=0.95\columnwidth]{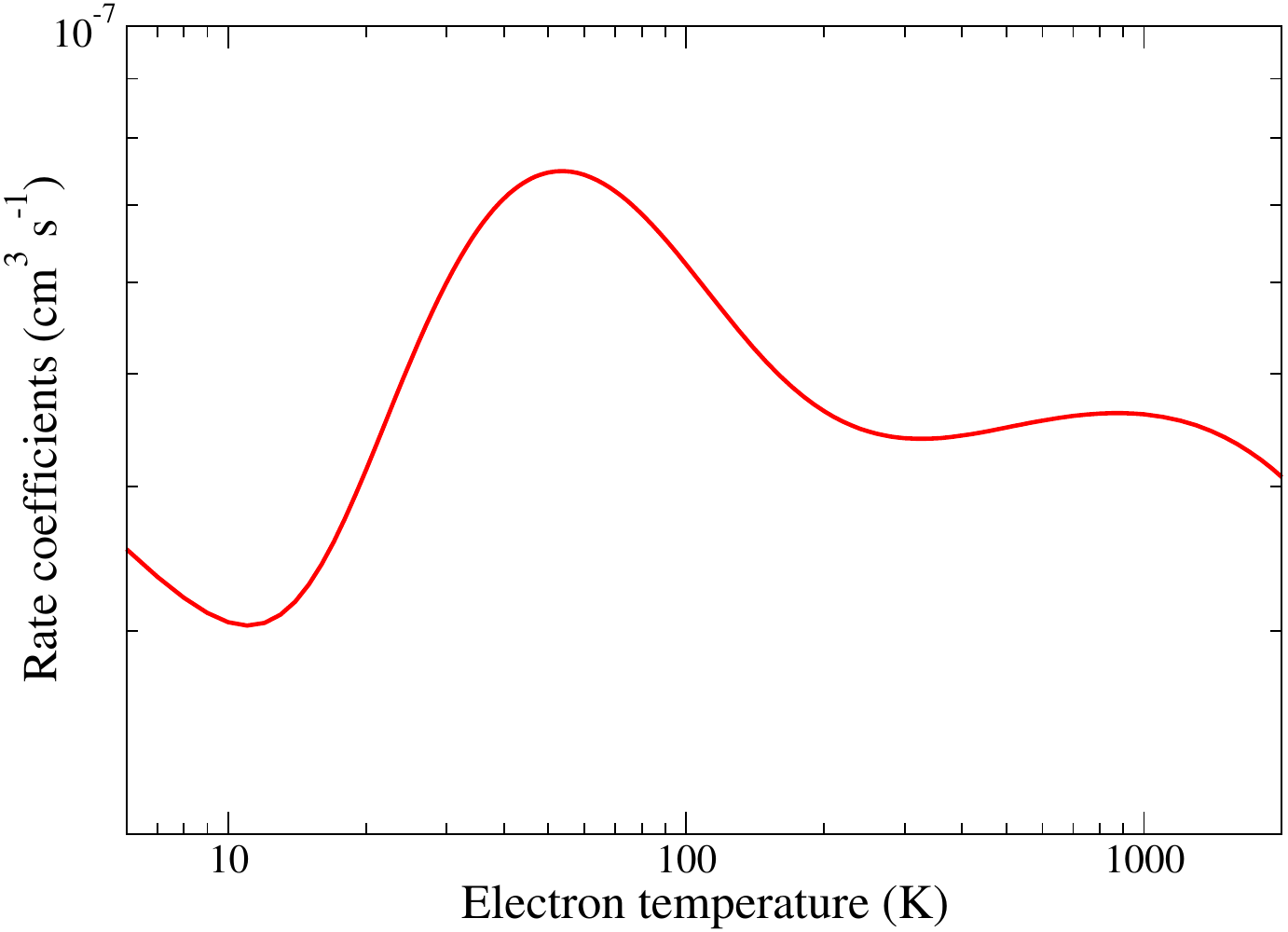}
\caption{
\label{c123rate}
(Color online) CH$^+$  DR rate coefficients after averaging with anisotropic Maxwell distribution given in Ref. \cite{amitay1996}. The result includes all three ion cores.
}
\end{center}
\end{figure}

\subsection{Rate coefficients}

The total DR cross sections considering three cores have been converted to isotropic (thermal) Maxwell rate coefficients up to $2000$ K. The rate coefficient is displayed in Figure \ref{c123rate} and shows broad peaks around $50$ K and $1000$ K clearly due to the presence of excited cores as can be seen by comparing with Figures \ref{c123conv2} and \ref{c123rate}.

\section{Conclusion}

We have demonstrated the role of core excited Rydberg states of CH in the dissociative recombination process~\cite{carata2000}. Our cross sections (first and second order) agree much better with the available experiments than the first order calculations previously undertaken by Carata {\it et al.} \cite{carata2000} in the energy range considered. Our present approach being validated by the comparison with accurate measurements, new computations extending to excited vibrational states and to competitive processes - elastic and inelastic collisions - will be initiated, in order to provide data for collisional-radiative modelling of the kinetics of interstellar molecular clouds and fusion edge plasmas in the low-temperature regions.

The results presented here are of fundamental importance for the modelling of all cold ionized media, and very promising, suggesting the need for further studies, especially with improved and more accurate potential energy curves and couplings, not just for the molecular ion, but for the neutral molecule as well.

Whereas the present approach takes into account the full vibrational structure of the existing relevant electronic states of CH$^+$ and CH, assuming complete rotational relaxation, further work will be necessary in order to take into account the role of rotational effects (excitation and inter-channel couplings) on the magnitude of the rate coefficient at very low energy.

\section*{Acknowledgements}
The authors acknowledge support from the IAEA via the Coordinated Research Project ``Light Element Atom, Molecule and Radical Behaviour in the Divertor and Edge Plasma Regions", from Agence Nationale de la Recherche via the projects `SUMOSTAI' (ANR-09-BLAN-020901) and  `HYDRIDES' (ANR-12-BS05-0011-01), from the IFRAF-Triangle de la Physique via the project `SpecoRyd', and from the CNRS via the programs `Physique et Chimie du Milieu Interstellaire', and the  PEPS projects `Physique th\'{e}orique et ses interfaces' TheMS and TPCECAM. They also thank for generous financial support from La R\'egion Haute-Normandie via  the GRR Electronique, Energie et Mat\'eriaux, from the ``F\'ed\'eration de Recherche Energie, Propulsion, Environnement'', and from the LabEx EMC$^3$ and FEDER via the projects PicoLIBS (ANR-10-LABEX-09-01), EMOPlaF and CO$_2$-VIRIDIS. IFS also thanks the Laboratoire  Aim\'e Cotton for hospitality. JZM and KH acknowledge support from USPC via ENUMPP and Labex SEAM. KC thanks the department INSIS of CNRS for a research grant in 2013, and LOMC for hospitality.

\section*{Data availability}
Upon a reasonable request, the data supporting this article will be provided by the corresponding author.

\end{document}